**Analytic Harmonic Tunable Closed Orbits of Trapped N-Body Systems**

Joseph West
Department of Chemistry and Physics
Indiana State University

**Abstract**
The motion of each particle in an N-body system of identical masses interacting via an attractive **or repulsive** pair-wise linear force law (the "Swarm") **and** with an external attractive **or** repulsive linear force law (the "Trap") is considered. In all Swarm-Trap combinations the motion of all N particles is completely separable and positions are found as a function of time in simple analytic form. In all attractive Traps the center of mass of the Swarm is bound to the center of the trap. For attractive or weakly **repulsive** Swarms in attractive Traps the particles are bound to the center of mass of the Swarm and an infinite set of Swarm/Trap force constant ratios result in every particle executing a **closed periodic orbit**. Figures are provided of trajectories for various combinations of Swarm and Trap force constants. The derivations are suitable for the advanced undergraduate classroom.

**1. Introduction**

Topics of simple harmonic motion (SHO), orbital motion in various power laws, and the dynamics of N-body systems have long been of interest in research and the advanced undergraduate classroom. As shown by Bertrand's theorem,[1] the linear and inverse square force laws are the only central power laws that lead to closed orbits for a range of arbitrary initial conditions. A system of N particles in a "free Swarm" that interact via the pair-wise linear force law has proven to be particularly accommodating in these regards. The force law is given by

$$\mathbf{F}_{Jn} = -J m_n m_j (\vec{r}_n - \vec{r}_j) \qquad (1)$$

where the force on particle $m_n$ is directed along line connecting the particles, $\mathbf{r}_n$ and $\mathbf{r}_j$ are the position vectors and J is a constant with units of N/(kg²m). In an Attractive Swarm (J > 0) every particle moves in a closed, harmonic, elliptical orbit with all N particles having an identical orbital period. For a Repulsive Swarm (J < 0), each particle increases its distance from the origin exponentially in time with an identical growth factor.[2] In addition, it has been known for some time that the equivalent quantum N-body system of **identical** particles $\{m_n = m\}$ is completely separable, giving simple products of SHO wavefunctions as the system wavefunction.[3,4]

An "optical lattice" can provide an equivalent of a 3D harmonic potential for particle trapping and in optical computing.[5] Ions are also routinely studied in Paul Traps, which provide a (nearly) ideal quadrupole potential produced by a cylinder of charge and oppositely charged endcaps. The potential does not have a stable minimum in that case, but it is stabilized via an additional logarithmic radial potential provided by a charged wire along the central axis of the trap.[6-9]

In this paper the particles of an N-body "Swarm" **are restricted to all having the same mass**. This is more restrictive than Ref 2, but most particle trapping applications are focused on collections of identical masses. It is shown that attractive **and** repulsive Swarms in attractive **and** repulsive linear force law "Traps" allow for separable, analytic, closed form solutions for the position of every particle as a function of time for arbitrary initial conditions. For suitable choices of the Swarm and Trap force constants, all N particles move in **closed orbits**. In the case of Repulsive Swarms in Attractive Traps, it is possible to "tune" the qualitative behavior of the entire Swarm between a stable orbit regime, and an exponentially exploding trajectory regime by an adjustment of the Trap strength, or the number of particles in the trap.

As a speculative application one could consider simple geometric constructs of uniform density material as Dark Matter "Traps." The following mass/charge distributions produce SHO potentials within the objects, even for Dark Matter particles:[10,11]

Uniform Slab:      1D along axis perpendicular to the surface
Uniform Cylinder:  2D in planes perpendicular to the axis of the cylinder
Uniform Sphere:    3D radially from the center of the sphere

The linear force constant of each trap is directly proportional to the density of the material used for fabrication. Even if dark matter particles were mutually repulsive, they could move in closed elliptical orbits in these simple "gravity traps." Dark matter could form its own harmonic attractive spherical gravitational trap, if it is present in a roughly uniform dense spherical distribution. The low density of common materials make it unlikely such traps would be useful at the laboratory scale, but neutron star interiors or stellar cores could be potential candidates.[12]

The remainder of the paper is organized as follows: section 2 considers the general form of the equations of motion and the derivation of the motion of the center of mass is presented; in section 3 the equations of motion for the individual masses in the Swarm are solved; and concluding remarks are provided in section 4.

## 2. Center of Mass Motion

Consider placing a Swarm of N identical particles interacting with the mutual interactions of Eq. (1) into a linear force law "Trap" of the form

$$\mathbf{F}_{Tn} = -K_x x \hat{e}_x - K_y y \hat{e}_y - K_z z \hat{e}_z \qquad (2)$$

The values of $K_x$, $K_y$, and $K_z$ determine whether each force component of the Trap is directed towards (an "Attractive Trap," $K > 0$), or away from (a "Repulsive Trap," $K < 0$) the origin. An electrostatic trap in free space must meet the Laplace condition: $K_x + K_y + K_z = 0$, the most typical application being an ideal quadrupole trap ($K_x = K_y = -K_z/2$).[6-9] On the other hand one expects optical lattice traps to be isotropic ($K_x = K_x = K_z = K$). The derivations follow exactly the same for both cases, and are identical for 1D – 3D cases so only the x-component derivations are presented.

Applying Newton's Second Law to "particle n" gives

$$m\ddot{x}_n = -K x_n - Jmm \sum_L (x_n - x_L) \qquad (3)$$

$$\ddot{x}_n = -\frac{K x_n}{m} - Jm \sum_L (x_n - x_L) \qquad (4)$$

$$\ddot{x}_n = -\left(\frac{K}{m} + JM\right) x_n + JMR_x \equiv -\left(\frac{K}{m} + SN\right) x_n + SNR_x \qquad (5)$$

where the summations are from L = 1 up to L = N, and a "dot" indicates a derivative with respect to time. The position, velocity and acceleration of the center of mass is found by summing over the N particles

$$R_x \equiv \frac{\sum_n m_n x_n}{M} = \frac{\sum_n x_n}{N} \qquad (6)$$
$$M\dot{R}_x = \sum_n m\dot{x}_n \qquad (7)$$
$$M\ddot{R}_x = \sum_n m\ddot{x}_n \qquad (8)$$

Substituting from Eq. (5) into Eq. (8) gives

$$M\ddot{R}_x = -K \sum_n x_n - Jmm \sum_n \sum_L (x_n - x_L) \qquad (9)$$
$$M\ddot{R}_x = -K \sum_n x_n - SNm[\sum_n(x_n) - \sum_L(x_L)] = -K \sum_n x_n \qquad (10)$$

The last two sums in Eq. (10) are identical leaving the well-known SHO and exponential forces/solutions for R(t)

$$\ddot{R}_x = -(K/m)R_x \qquad W^2 \equiv \left|\frac{K}{m}\right| \qquad (11)$$
$$R_x(t) = B_x \sin(Wt) + D_x \cos(Wt) \qquad K > 0 \qquad (12)$$
$$R_x(t) = B_x e^{Wt} + D_x e^{-Wt} \qquad K < 0 \qquad (13)$$

where the constants $B_x$ and $D_x$ are determined from the initial conditions ($\{x_n(0)$, and $\{vx_n(0)\}$) via Eqs. (6) and (7).

## 3. Swarms in Traps

The effect of $R_x(t)$ from Eqs. (12) and (13) on the Swarm is mathematically the same as harmonic or exponential forcing of a simple harmonic oscillator.[14,15] This allows a direct path to determine $\{x_n(t)\}$ for all of the Swarm-in-Trap combinations. In all of the example trajectories provided in the figures, the Trap is isotropic ($K_x = K_y = K_z = K$), the initial positions and velocities were generated by the built-in Rand() function of Excel, except where noted, and N = 4.

### 3.1. Attractive Swarm : Attractive Trap (J > 0, K > 0)

The first case is the one most easily visualized. After substituting $R_x(t)$ from Eq. (12) into Eq. (5), the solution is found by simple guess and substitution

$$\ddot{x}_n = -(W^2 + w^2)x_n + w^2 R_x(t) \qquad (14)$$
$$x_n(t) = b_{xn}\sin(\Omega t) + d_{xn}\cos(\Omega t) - \left(\frac{w^2}{W^2}\right)[B_x\sin(Wt) + D_x\cos(Wt)] \qquad (15)$$
$$w^2 \equiv |SN| \qquad \Omega^2 \equiv W^2 + w^2 \qquad (16)$$

where the constants $\{b_{xn}, d_{xn}\}$ are determined by the initial positions and velocities of the individual particles. When considered in 3D, the particles move in closed orbits **as viewed from the center of mass**, while the center of mass moves in a closed elliptical orbit about the origin. "Generic" example 2D trajectories are provided in Figs. 1 and 2. For both figures for w = 1 rad/s and W = 2 rad/s. The x(t) and y(t) positions of two particles and the center of mass are shown in Fig. 1, while the trajectory of one particle and the center of mass are shown in Fig. 2.

For Fig. 3 w = 12 rad/s and W = 3 rad/s, and the initial conditions are chosen so that |**R**| is "large" compared to that of the inter-particle spacing (|**R**-**r**$_n$|). This illustrates the fact that the individual particles do orbit about the center of mass of the Swarm. As already noted, when (w, W) are chosen as two legs of a Pythagorean triple, all N particles will move in closed orbits about the center of the Trap. This is illustrated in Fig. 4, with the values of w = 3 and W = 4 are chosen as two legs of the Pythagorean triple (w² + W² = 5²).

### 3.2. Repulsive Swarm : Repulsive Trap (J < 0, K < 0)

In all cases of a Repulsive Swarm in a Repulsive Trap (J < 0, K < 0) the center of mass moves to infinity (B$_x$ not zero) and the particles move infinitely far away from the center of mass (b$_{xn}$ not zero)

$$\ddot{x}_n = (W^2 + w^2)x_n + w^2 R(t) = \Omega^2 x_n + w^2 R(t) \qquad (17)$$
$$x_n(t) = b_{xn}e^{\Omega t} + d_{xn}e^{-\Omega t} + \left(\frac{w^2}{W^2}\right)[B_x e^{Wt} + D_x e^{-Wt}] \qquad K<0, J>0 \qquad (18)$$

The overall growth rate of the Swarm in the Trap is faster than that of the same Swarm if it is "free" ($\Omega^2 > w^2$). The motion of the particles is as if N were increased and the center of mass were fixed at the center of the Trap. No figures of this motion are provided, as they do not seem to provide much insight.

### 3.3. Repulsive Swarm : Attractive Trap (J < 0, K > 0)

A Repulsive Swarm in an Attractive Trap allows two different qualitative behaviors: bound stable orbits or exponentially exploding motion. This is **determined by whether K/SN > 1 or < 1.** Substituting R(t) from Eq. (12) gives

$$x_n(t) = b_{xn}\sin(\gamma t) + d_{xn}\cos(\gamma t) + B_x\sin(Wt) + D_x\cos(Wt) \qquad |K|>SN \qquad (19)$$
$$x_n(t) = b_{xn}e^{\gamma t} + d_{xn}e^{-\gamma t} - \left(\frac{w^2}{W^2}\right)[B_x\sin(Wt) + D_x\cos(Wt)] \qquad |K|<SN \qquad (20)$$
$$\gamma^2 \equiv |W^2 - w^2| \qquad (21)$$

In a strong Trap (Eq. (19)) the N mutually **repulsive** particles move **in stable orbits**. For w and W representing any two leges of a Pythagorean triple those are closed periodic orbits about the origin. The plots in Fig. 5 show bound trajectories about the center of mass for w < W. Figure 6 shows the closed orbital trajectory of particle n = 1 for (w = 3 rad/s, W = 5 rad/s), the Pythagorean triple (W² – w² = 4²).

In a "weak" Trap (|K| < SN, Eq. (20)), **all N particles** escape to infinity, while the center of mass executes a closed elliptical orbit about the origin. The exponential growth rate of the N particles is slower than that of a free Repulsive Swarm ($\gamma^2 < w^2$). This is illustrated in Fig. 7.

### 3.4. Attractive Swarm : Repulsive Trap (J > 0, K < 0)

An Attractive Swarm in a Repulsive Trap also allows two different qualitative behaviors: bound stable orbits, or exponentially exploding motion. This is **determined by whether K/SN > 1 or < 1.** Substituting R(t) from Eq. (13) gives

$$x_n(t) = b_{xn}\sin(\gamma t) + d_{xn}\cos(\gamma t) + \left(\frac{w^2}{W^2}\right)[B_x e^{Wt} + D_x e^{-Wt}] \qquad |K|<SN \qquad (22)$$

$$x_n(t) = b_{xn}e^{\gamma t} + d_{xn}e^{-\gamma t} + \left(\frac{w^2}{W^2}\right)[B_x e^{Wt} + D_x e^{-Wt}] \qquad |K| > SN \qquad (23)$$

In a "weak" Trap ($|K| < SN$, Eq. (22)), all N particles orbit about the center of mass as the center of mass moves exponentially away from the origin ($B_x$ not zero). In a "strong" Trap ($|K| > SN$, Eq. (23)), the particles move **away from the center of mass** ($b_{xn}$ not zero) exponentially, as the center of mass moves away from the origin.

### 3.5. Quadrupole Trap

The idealized quadrupole potential is a central component of the theoretical model of Penning and Kingdon ion traps.[6-9] The solution for this non-isotropic Trap are therefore of particular interest. Only Repulsive Swarms need to be considered because the trapped ions share the same net charge and so are mutually repulsive. The Trap potential and force are given by

$$U(x,y,z) = U(r,z) = K\left(-\frac{r^2}{2} + z^2\right) \qquad (24)$$
$$\mathbf{F}_{Tn} = Kx\hat{e}_x + Ky\hat{e}_y - 2Kz\hat{e}_z \qquad (25)$$

Conveniently, the solutions are given in terms of those already presented. The relevant trajectories are given by the following expressions

**Center of Mass**
$$R_x(t) = B_x \sin(Wt) + D_x \cos(Wt) \qquad (26)$$
$$R_y(t) = B_y \sin(Wt) + B_x \cos(Wt) \qquad (27)$$
$$R_z(t) = B_z e^{Wt\sqrt{2}} + D_z e^{-Wt\sqrt{2}} \qquad (28)$$

**Particles in a Weak Trap**
$$x_n(t) = b_{xn}e^{\gamma t} + d_{xn}e^{-\gamma t} - \left(\frac{w^2}{W^2}\right)[B_x \sin(Wt) + D_x \cos(Wt)] \qquad (29)$$
$$y_n(t) = b_{yn}e^{\gamma t} + d_{tn}e^{-\gamma t} - \left(\frac{w^2}{W^2}\right)[B_y \sin(Wt) + D_y \cos(Wt)] \qquad (30)$$
$$z_n = b_{zn}e^{\Omega t} + d_{zn}e^{-\Omega t} + \left(\frac{w^2}{W^2}\right)[B_z e^{Wt} + D_z e^{-Wt}] \qquad (31)$$

**Particles in a Strong Trap**
$$x_n(t) = b_{xn}\sin(\gamma t) + d_{xn}\cos(\gamma t) + B_x \sin(Wt) + D_x \cos(Wt) \qquad (32)$$
$$y_n(t) = b_{yn}\sin(\gamma t) + d_{yn}\cos(\gamma t) + B_y \sin(Wt) + D_y \cos(Wt) \qquad (33)$$
$$z_n = b_{zn}e^{\Omega t} + d_{zn}e^{-\Omega t} + \left(\frac{w^2}{W^2}\right)[B_z e^{Wt} + D_z e^{-Wt}] \qquad (34)$$

All constants are determined by the initial positions and velocities of the N particles $\{\mathbf{r}_n(t), \mathbf{v}_n(t)\}$.

### 4. Conclusions

A system of N particles interactive pairwise via a linear force law (a Swarm) that is interacting with an external linear force law (a Trap) allows for a complete simple analytic solutions of the motion of every particle as a function of time. Such Trap and N-body systems are of interest in applications as model systems for mass spectroscopy and quantum computing. Strong attractive traps are found to produce closed orbits with identical periods for all N particles when the ratio of the strengths of the trap (K) and Swarm interactions (JM) are chosen as two legs of a Pythagorean triple. Almost any classical N-body system providing analytical solutions of position and velocity explicitly as functions of time is of interest.[16-19] The derivations and expressions presented here are of interest and useful in the graduate or advanced undergraduate classroom.

    Future work with Trapped Swarms will examine Pressure-Volume-Temperature relationships;[2] determine the feasibility of the uniform density dark matter "gravity traps;" a as a toy model of dark matter particle motion in uniform density particle traps; examine the self-trapping of dark matter in spherical distributions; and consider the interactions of Swarms with externally applied ac signals near resonance with w or W, specifically considering gravity waves as an external excitation of Dark Matter motion in density gravity traps.

**Figures**

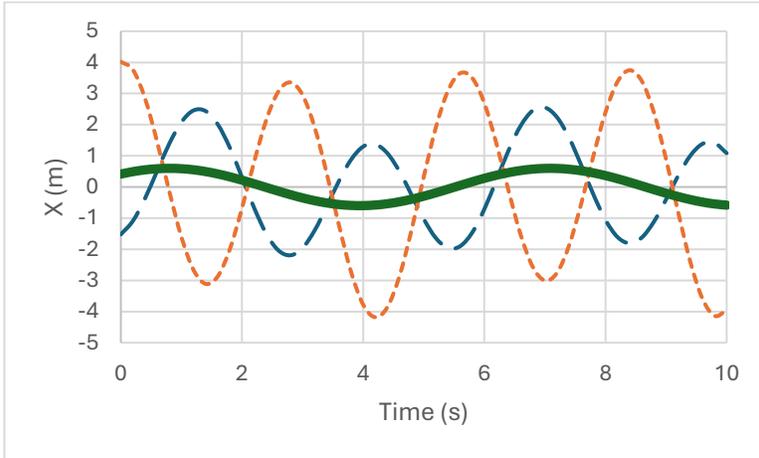

(a)

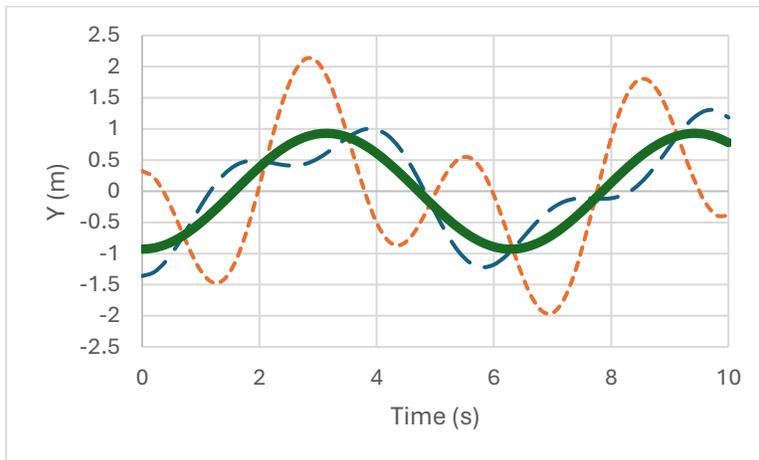

(b)

Figure 1. An Attractive Swarm in an Attractive Trap (w = 1 rad/s and W = 2 rad/s) showing time dependence of (a) x-position (b) y-position for particle 1 (short dashes) particle 2 (long dashes) and the center of mass (solid line).

(a)
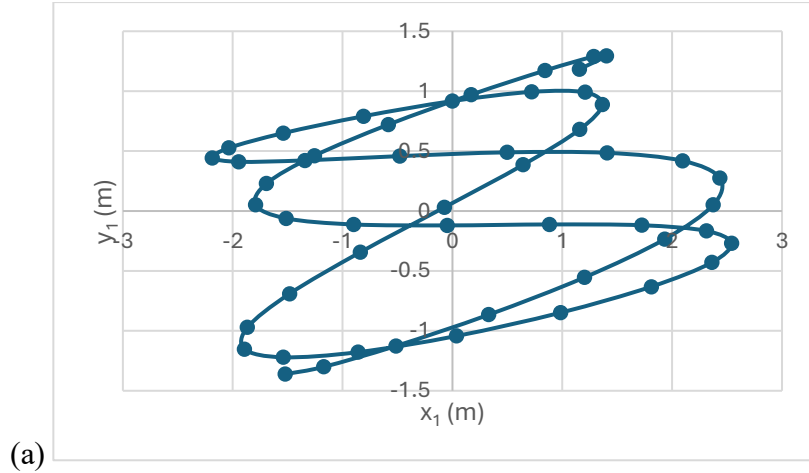

(b)
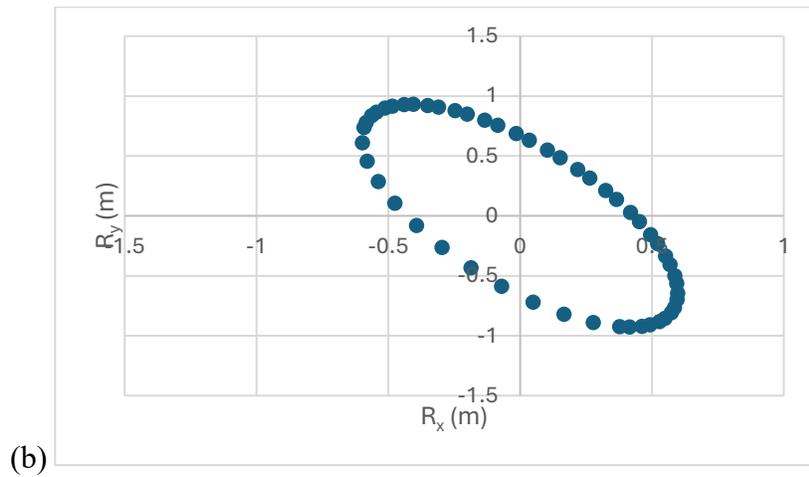

Figure 2. The same parameters as shown in Figure 1, showing the trajectory of particle 1 (a) and the center of mass (b).

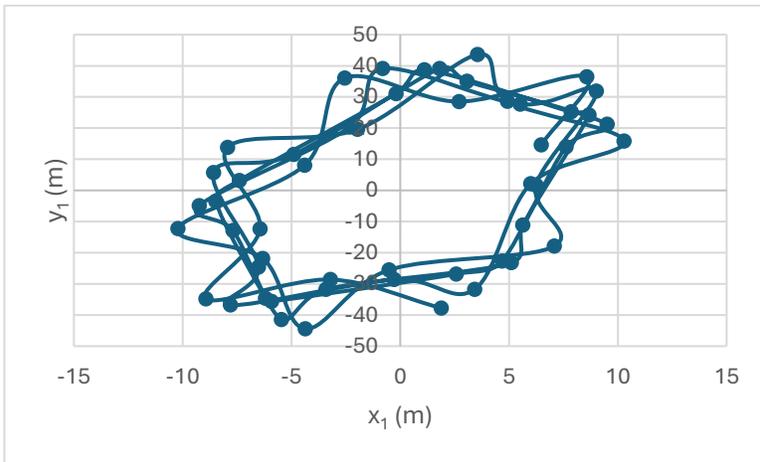

(a)

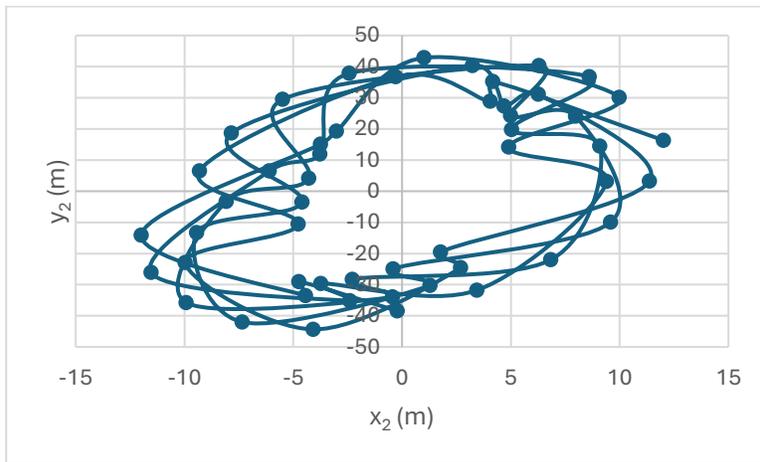

(b)

Figure 3. An Attractive Swarm in an Attractive Trap (w = 12 rad/s, W = 3 rad/s). The initial conditions were intentionally chosen to produce a "slow large" radius orbit of the center of mass, with "fast small" Swarm motion about the center of mass. In (a) the trajectory of particle n = 1 and in (b) the trajectory of particle n = 2.

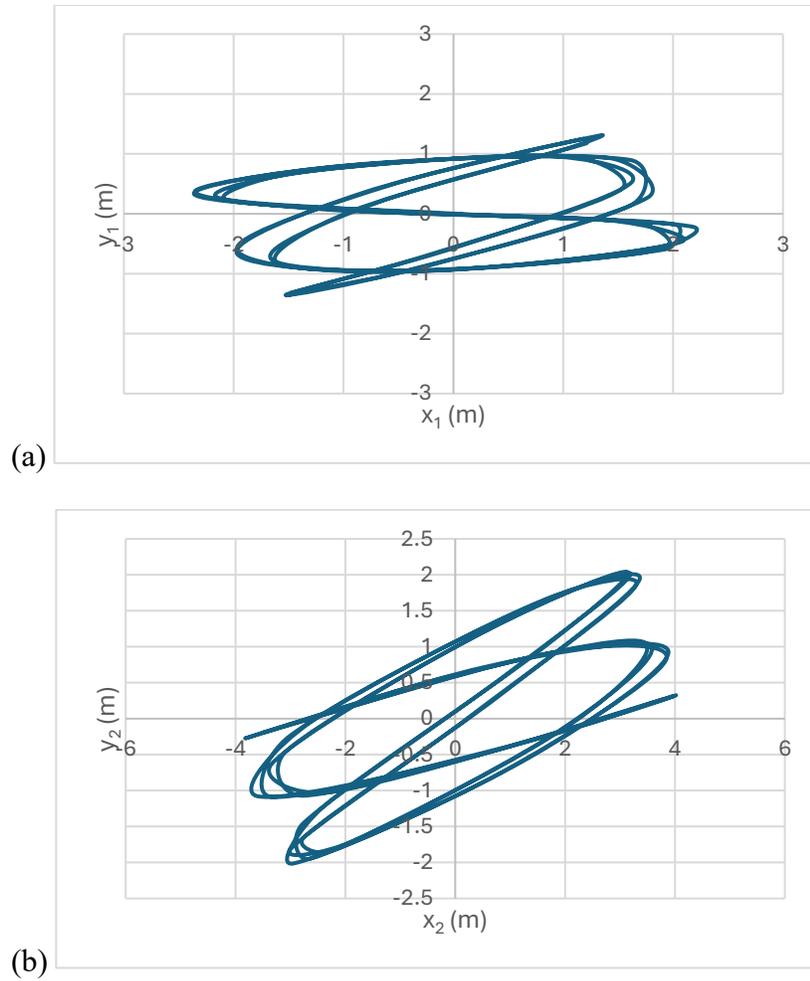

(a)

(b)

Figure 4. An Attractive Swarm in an Attractive Trap (w = 3 rad/s, W = 4 rad/s) with w, W chosen as two legs of a Pythagorean Triple. The center of mass and all N particles follow closed periodic orbits with period (T = 2pi/(sqrt(w²+W²)) = 1.26 s). The trajectories of particle n = 1 (a) and particle n = 2 (b) are closed periodic orbits in the Trap frame of reference.

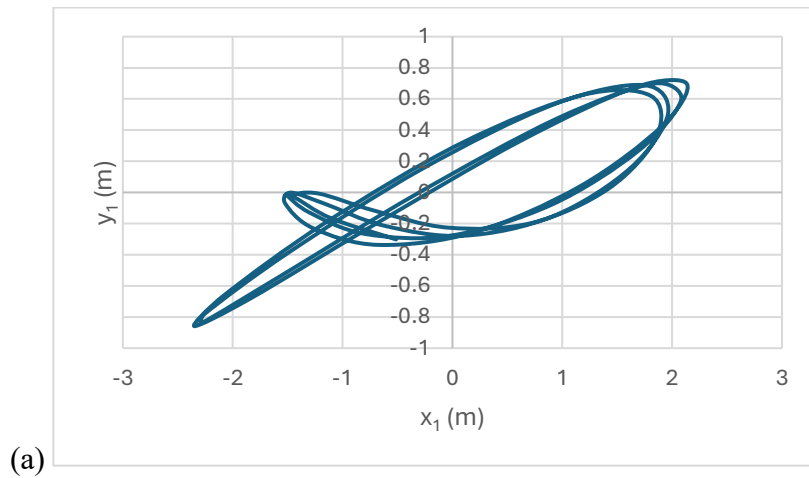

(a)

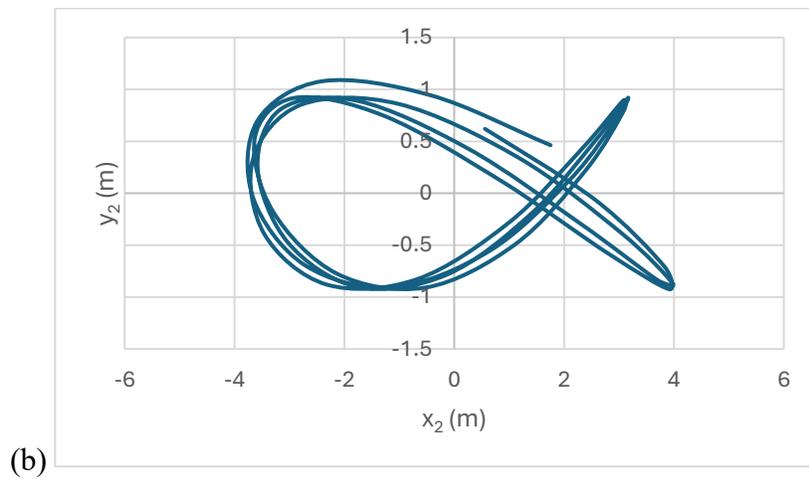

(b)

Figure 5. A Repulsive Swarm in a Strong Attractive Trap (K > |JM|, w = 3 rad/s, W = 4 rad/s), specifically chosen as a case that is **not** a Pythagorean triple. The trajectories of particle n = 1 (a) and particle n = 2 (b) are shown.

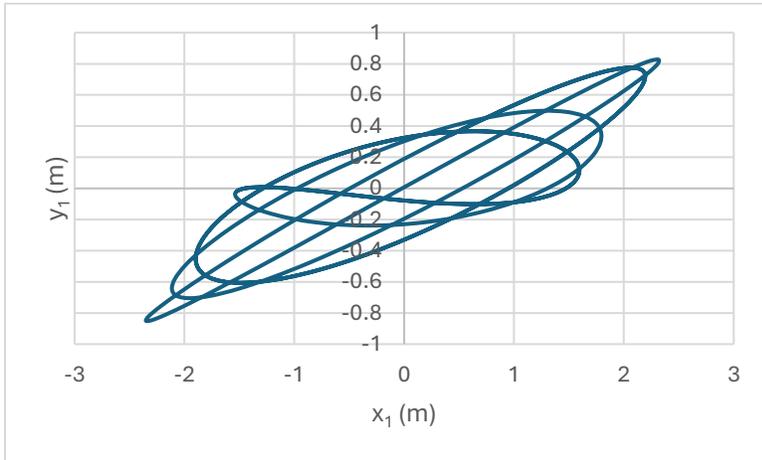

Figure 6. A Repulsive Swarm in a Strong Attractive Trap (K > |JM|, w = 3 rad/s, W = 5 rad/s), specifically chosen so that it is a Pythagorean Triple. The trajectory of particle (n = 1) is shown for 3 full periods. All N particles move in similarly closed orbits.

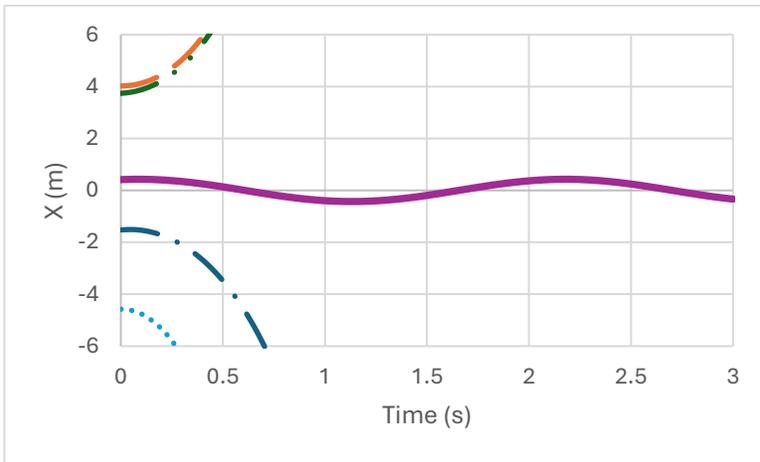

(a)

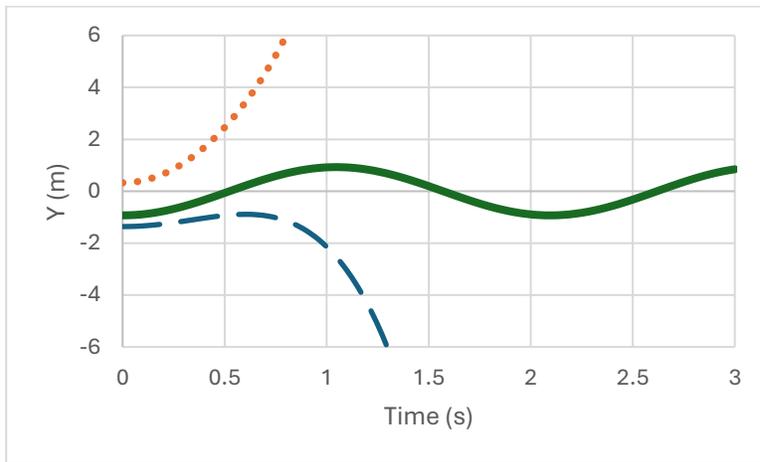

(b)

Figure 7. A Strong Repulsive Swarm in a Weak Attractive Trap ($|JM| > K$, w = 4 rad/s, W = 3 rad/s) with w, W chosen as two legs of a Pythagorean triple. The resulting exponential expansion of all N particles is not a noticeably different than that of a Repulsive Free Swarm. In (a) the x component motion of all four masses, as well as that of the center of mass (heavy solid oscillating line) are shown ($x_2$ and $x_4$ are close together in the upper left corner). In (b) the y component motion is shown for $x_1$ (long dashes) and $x_2$ (short dashes) and the center of mass (heavy solid line).